\documentclass[superscriptaddress,preprint]{revtex4-1}
\pdfoutput=1
\usepackage{hyperref}
\usepackage{graphicx, subfigure, amssymb, bm}
\usepackage[usenames, dvipsnames]{color}
\usepackage{amsmath, mathrsfs}
\usepackage{physics}

\newcommand{\RomanN}[1]{\uppercase\expandafter{\romannumeral #1}}
\newcommand{\romanN}[1]{\expandafter{\romannumeral #1}}
\allowdisplaybreaks[4]
\begin{document}

\title{Remote weak signal measurement via bound states in optomechanical
system}
\author{Xun Li}
\affiliation{School of Physics, Dalian University of technology,
Dalian 116026, P.R.China}
\affiliation{National Key Laboratory of Shock Wave and Detonation
Physics, Institute of Fluid Physics, China Academy of Engineering
Physics, Mianyang 621900, China}
\author{Biao Xiong}
\affiliation{School of Physics, Dalian University of technology,
Dalian 116026, P.R.China}
\author{Shilei Chao}
\affiliation{School of Physics, Dalian University of technology,
Dalian 116026, P.R.China}
\author{Chengsong Zhao}
\affiliation{School of Physics, Dalian University of technology,
Dalian 116026, P.R.China}
\author{Hua-Tang Tan}
\email{tht@mail.ccnu.edu.cn}
\affiliation{Department of Physics, Huazhong Normal University, Wuhan
430079, China}
\author{Ling Zhou}
\email{zhlhxn@dlut.edu.cn}
\affiliation{School of Physics, Dalian University of technology,
Dalian 116026, P.R.China}

\begin{abstract}
A scheme for remote weak signal sensor is proposed in which a  coupled
resonator optical waveguide~(CROW), as a transmitter, couples to a hybrid optomechanical
cavity and an observing cavity, respectively. The non-Markovian
theory is employed to study the weak force sensor by treating the CROW as a
non-Markovian reservoir of the cavity fields,
and the negative-effective-mass~(NEM) oscillator is introduced to
cancel the back-action noise.
Under certain conditions, dissipationless bound states can be formed
such that weak signal can be transferred in
the CROW without dissipation. Our results show that ultrahigh sensitivity
can be achieved with the assistance of the bound states under certain parameters regime.

\noindent \textbf{Keywords:} cavity optomechanics, remote detection,
non-Markovian environment, quantum noise

\end{abstract}

\maketitle

\section{Introduction}

Optomechanical system, involving the coupling between mechanical
motion and cavity field, provides us a high sensitive device to detect
weak force, tiny mass and displacement of the mechanical
motion~\cite{Aspelmeyer2014a,Chen2013}. With the advance of micro-nano
technology, micro-cavity optomechanical systems with high mechanical
frequency, high quality factor and strong optomechanical coupling are
realized in several kinds of systems such as whispering-gallery-mode
resonator~\cite{Cai2000,Tomes2009,hao2018,Schliesser2009,Shen2016}, levitated
nano-sphere~\cite{Reimann2018a,Hoang2016a} and optomechanical
crystal~\cite{Eichenfield2009,Bochmann2013}. These progresses push
optomechanical systems in the precision detection further into
application. Approaches to force
detection~\cite{Gavartin2012,Doolin2014} based on nano-mechanical
systems are well-established and have been used for measuring
displacement~\cite{Anetsberger2009}. Increasingly, it is believed that
next-generation mechanical biosensor may be realized in
nano-mechanical systems, because they are particularly matched in
size with molecular interactions, and provide a basis for biological
probes with single-molecule sensitivity~\cite{Erdil2019,Arlett2011}.
For the biosensing and medical diagnoses or other detection scenario,
the local detection scheme might not meet the needs of practical
demand,
it is necessary to construct waveguide-optomechanical
coupling system so as to perform remote detection.
To our knowledge, this remote force detection has not yet been
investigated.

The sensitivity of optomechanical detector is limited by the noise.
Various proposals have been put forward for reducing noise, including
squeezing of mechanical
oscillator~\cite{Lu2015,Lei2016a,Wang2014,Zhang2019}
frequency-dependent squeezing of optical
field~\cite{Bondurant1984,ZhaoWen2020,Ma2014}, two-tone measurements, dual
mechanical oscillator configurations~\cite{Briant2003,Woolley2013},
and atomic assistance
detection~\cite{Motazedifard2016,Bariani2015,Tesfay2020}.
Especially, it has been shown that quantum back-action~(QBA)
noise can be cancelled when the prob field couples to positive and
negative-effective-mass~(NEM) oscillators
simultaneously~\cite{Tsang2010,Zhang2013a,Motazedifard2016}. The QBA
free proposal has been realized in a hybrid cavity optomechanical system
in which a spin ensemble plays the role of the negative-mass
oscillator~\cite{M2017Quantum}.\\
\indent For remote detection, the waveguide is usually employed to
connect the sensor and the detector. Using tapered fibers coupling to
sensing cavity has been investigated in~\cite{Chen2017,Anetsberger2009}. In the purpose
of integrating the system on chip and improving the detection precision, the
waveguide integrated in the sensing cavity has been realized in microcavity
regime~\cite{Liu2018,wang2010}. Theoretically, the waveguide can be treated
as structured reservoirs~\cite{DeVega2017,Longhi2006,Lodahl2015,Liao2010a,Xu2017}, and the
theory of non-Markovian quantum open system is an effective method to
study the dynamics of the objects coupling to the reservoir. In cavity
quantum electrodynamics regime, the structured reservoir can be photonic
crystals or waveguides~\cite{Hoeppe2012,Tan2011,Ciccarello2015,Ballestero2013,Lodahl2015,PhysRevA.99.032101}. It has been
shown that the bound states without dissipation can be formed when system
coupled to band gaps or finite band spectrum~\cite{Zhang2012,Longhi2006}
which is easily satisfied in photonic crystals or waveguides~\cite{DeVega2017,Tan2011,Longhi2006,Quang1997,Hsu2016}. The dissipationless of
bound state benefits the transfer of the signals.\\
\indent Since the QBA evading in hybrid optomechanical system had been realized in experiment ~\cite{M2017Quantum}, in the paper, we put forward a proposal by generating QBA evading measurement to remote force detection.  Using non-Markovian theory, we solve the dynamics and
obtain the output signal of the hybrid system. We carefully investigate the
surviving condition of the bound state and show that the output fields can
be transferred in the presence of bound state. A high precision and
minimized weak force sensor can be achieved.
Different from the researches~\cite{Gavartin2012,Doolin2014,Zhang2016b}, we
consider a remote weak force detection, which may be more suitable in
some cases. In order to avoid the photon consumption of the waveguide, we
investigated the condition of the bound state, which should be meaningful for
experimental realization.

This paper is arranged as follows.
In Sec.~\ref{sec:Moldel-Hamiltonian}, we present the model and
Hamiltonian of our proposal.
We study the effective non-Markovian reservoir and the bound states in
Sec.~\ref{sec:the_dynamics_of_system}.
The sensitivity and the mechanism of suppressing the noise are
discussed in Sec.~\ref{sec:sensitivity-weakSignal}.
Finally, Sec.~\ref{sec:discussion_and_conclusion} gives a summary of
this work.
\section{Model and Hamiltonian}
\label{sec:Moldel-Hamiltonian}

\begin{figure}[tbph]
\centering\includegraphics[width=0.9\linewidth]{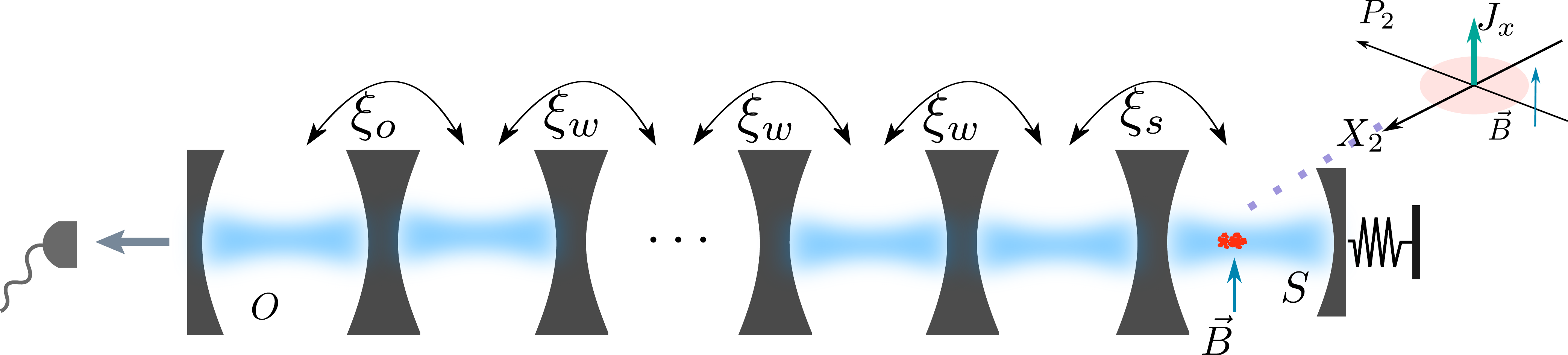}
\caption{The sketch of hybrid optomechanical system where a waveguide
couples to the optomechanical cavity $S$ and an observing cavity $O$. The atomic
ensemble in a static magnetic field with specific direction can
be regarded as a NEM spin oscillator (detail structure see ~\cite{M2017Quantum}).}
\label{fig:sketch}
\end{figure}
In order to detect remote weak signal, we employ a hybrid cavity
optomechanical system, shown in Fig.~\ref{fig:sketch}. In the cavity
$S$, the movable mirror works as a sensor, and the fixed mirror of the
cavity $S$ is connected to a CROW so that the signal can be
transmitted to the observing cavity $O$ and be read out by homodyne
detector. In addition, an atomic ensemble whose spin direction
$\va{J}$ can be manipulated by a magnetic field $\va{B}$ is placed
in cavity $S$. Employing the same
procedure as in~\cite{M2017Quantum,Motazedifard2016}, the spin of the
atomic ensemble can be effectively equivalent to an oscillator with
effective-negative-mass;
therefore, the noise of quantum back-action can be evaded. The
Hamiltonian of the system can be written as
\begin{equation}
H=H_{OM}+H_{\text{crow}},
\end{equation}
with
\begin{equation}
\begin{split}
H_{OM}=& \frac{p_{1}^{2}-p_{2}^{2}}{2m}+\frac{m}{2}\omega_{m}^{2}\qty
(q_{1}^{2}-q_{2}^{2}) +\hbar ga_{s}^{\dagger }a_{s}(q_{1}+q_{2})
+\hbar \omega_{s}a_{s}^{\dagger }a_{s} 
+f q_{1}.
\end{split}
\end{equation}
The first and the second terms express the Hamiltonian of the mechanical
oscillator and the NEM oscillator (the detailed description of the atom ensemble
equivalence to NEM oscillator can be referred in~\cite{M2017Quantum})
where $q_{1}$~($p_{1}$) is the position~(momentum) operator of the
mechanical oscillator, $q_{2}$~($p_{2}$) is the position~(momentum)
operator of NEM oscillator.
To cancel the back-action noise, we let the two oscillators
have the same mass~($m$) and frequencies ($\omega_{m}$) as had reported by Ref.~\cite{M2017Quantum}.
The third term describe the coupling between the two
oscillators and the cavity $S$ where the NEM oscillator couples to the field
with the same form as the optomechanical interaction~\cite{Kohler2018}.
$a_{s}$~($a_{s}^{\dagger }$) is the
creation~(annihilation) operator of the sensing cavity $S$, and $g$ is the
coupling strength. 
The fourth term is the energy of the cavity $S$, and the last term
describes the weak force $f$ coupling to the
mechanical oscillator.

When defining collective position $Q=(q_{1}+q_{2})$ and collective momentum $P=
\frac{1}{2}(p_{1}+p_{2})$, relative position $\Phi =\frac{1}{2}(q_{1}-q_{2})$
and relative momentum $\Pi =p_{1}-p_{2}$, the Hamiltonian can be
transformed into
\begin{equation}
\begin{split}
H_{OM}=& \frac{P\Pi}{m}+m\omega_{m}^{2}Q\Phi +\hbar g a_{s}^{\dagger
}a_{s}Q+f(Q+2\Phi)+ \hbar \omega_{s}a_{s}^{\dagger}a_{s}.
\end{split}
\end{equation}
We have the commutation $[Q,\Pi]=0$ and $[P,\Phi]=0$.
Therefore, the collective position~$Q$~(momentum $P$) and relative
momentum~$\Pi$~(position $\Phi$) are a pair of observable operators
which can be simultaneously measured with arbitrary precision.

As shown in Fig.~\ref{fig:sketch}, the CROW consists of $N$-cavity
chain, and the CROW couple to the optomechanical sensing cavity $S$ and
observing cavity $O$, respectively.
The Hamiltonian of the CROW can be written as
\begin{equation}
\begin{split}
H_{\text{crow}}=& \sum_{n=1}^{N}\hbar \omega_{w}a_{n}^{\dagger
}a_{n}-\sum_{n=1}^{N-1}\hbar \xi_{w}(a_{n}^{\dagger }a_{n+1} +a_{n+1}^{\dagger }a_{n})+\hbar \xi_{s}(a_{1}^{\dagger
}a_{s}+a_{s}^{\dagger }a_{1}) \\
& \quad +\hbar \xi_{o}(a_{N}^{\dagger }a_{o}+a_{o}^{\dagger }a_{N})+\hbar
\omega_{o}a_{o}^{\dagger }a_{o}
+i\hbar E_{o}\qty(a_{o}^{\dagger }e^{-i\omega_{d}t}-a_{o}e^{i\omega_{d}t}),
\end{split}
\end{equation}
where the first term is the energy of CROW, the second term stands
for the hopping between the nearest neighbor cavity with rate $\xi_{w}$,
the third and the fourth terms describe the coupling of the cavity $S$ and the cavity $O$
to the $1$st and $N$th cavity of CROW respectively, where $a_{o}~(a^{ \dagger }_{o})$ is the annihilation and creation operator of cavity $O$, the fifth
term is the energy of cavity $O$, the last term represents the classical driven of cavity $O$ with frequency $\omega_{d}$ and strength $E_{o}$.
Performing the Fourier transformation~\cite{Tan2011}
\begin{equation}
a_{k}=\sqrt{\frac{2}{\pi }}\sum_{n=1}^{N}\sin (nk)a_{n}\qquad (0<k<\pi),
\end{equation}
then we can rewrite the Hamiltonian $H_{\text{crow}}$ as
\begin{equation}
\begin{split}
H_{\text{crow}} \big/\hbar=& \omega_{o}a_{o}^{\dagger }a_{o}+\sum_{k}\omega
_{k}a_{k}^{\dagger }a_{k} +\sum_{j=s,o} \sum_{k} V_{j}(k)(a_{_{j}}^{\dagger
}a_{k}+a_{k}^{\dagger }a_{_{j}})
+i E_{o}\qty(a_{o}^{\dagger }e^{-i\omega_{d}t}-a_{o}e^{i\omega_{d}t}),
\end{split}
\label{Eq:CROW-k}
\end{equation}
where $\omega_{k}=\omega_{w}-2\xi_{w}\cos k,$ $V_{j}(k)=\xi_{j}\sqrt{\frac{2}{\pi }}\sin (n_{j}k)$~($n_{s}=1,n_{o}=N$). The Hamiltonian Eq.~\eqref{Eq:CROW-k} implies that the CROW can be
regarded as a structured reservoir of the cavity $S$~$(O)$. In order
to exactly solve
the dynamics of the system, the non-Markovian treatment should be employed.
The coupled cavity chain or waveguide is equivalent to a structured reservoir
has been investigated in~\cite{Tan2011}. Here, we employ the coupled cavity
chain to transmit information of weak signal from the sensor to the
detector. For simplicity, we assume that the CROW is ideal without coupling to additional environment.

After switching into a frame rotating with respect to $H_{0}=\hbar\omega_{d}(a_{o}^{\dagger}a_{o} +a_{s}^{\dagger}a_{s}+\sum_{k}a_{k}^{\dagger}a_{k})$, and
nondimensionalizing operators with transform: $g \to g \sqrt{m\omega_m/\hbar}
$, $f \to f \sqrt{\hbar \omega_m m}$, $Q \to Q\sqrt{\hbar/m\omega_m}$, $\Phi
\to \Phi \sqrt{\hbar/\omega_m m}$, $P \to P \sqrt{\hbar \omega_m m}$ and $\Pi \to \Pi \sqrt{\hbar \omega_m m}$, the Hamiltonian can be changed into
time-independent form as
\begin{equation}
\begin{split}
H \big/\hbar=& \omega_m \qty(P\Pi+ Q\Phi)+ga_{s}^{\dagger}a_{s}Q+f(Q+2\Phi)
 +\Delta_{s}a_{s}^{\dagger}a_{s}+\Delta_{o}a_{o}^{\dagger}a_{o}+iE_{o}
\qty(a_{o}^{\dagger}-a_{o}) \\
& \; +\sum_{k}\Delta_{k}a_{k}^{\dagger
}a_{k}+\sum_{j=s,o} \sum_{k} V_{j}(k)(a_{j}^{\dagger}a_{k}+a_{k}^{\dagger}a_{j}),
\end{split}
\label{Eq:H-total-e}
\end{equation}
where $\Delta_{j}=\omega_{j}-\omega_{d}$~$(j=s,o,k)$. We will use
Eq.~\eqref{Eq:H-total-e} to calculate the output of the weak
signal~(weak force $f$).

\section{The effective non-Markovian reservoir and the bound states}

\label{sec:the_dynamics_of_system}

As we have pointed out that the coupled cavity chain is equivalent to a
structured reservoir, we now need to solve the dynamics with non-Markovian
theory.

Using the Hamiltonian Eq.~\eqref{Eq:H-total-e}, we can obtain the Heisenberg
equations as
\begin{subequations}
\label{Eq:Heisenberg-int}
\begin{align}
\dot{Q}=& \omega_{m}\Pi ,  \label{Eq:Heisenberg-Q} \\
\dot{\Pi}=& -\omega_{m} Q-2f-\frac{\gamma_{m}}{2}\Pi +\sqrt{\gamma_{m}}
\Pi ^{\text{in}},  \label{Eq:Heisenberg-Pi} \\
\dot{a}_{s}=& -i\Delta_{s}a_{s}-\frac{\kappa_{s}}{2}a_{s}-i g a_{s}Q-
i \sum_{k} V_s a_{k} +\sqrt{\kappa_{s}}a_{s}^{\text{in}},  \label{Eq:Heisenberg-aa}
\\
\dot{a}_{o}=& -i\Delta_{o}a_{o}-\frac{\kappa_{o}}{2}a_{o}-i\sum_{k} V_o a_{k}+E_{o} +\sqrt{\kappa_{o}}a_{o}^{\text{in}},  \label{Eq:Heisenberg-ab} \\
\dot{a}_{k}=& -i\Delta_{k}a_{k}-i\sum_{j=s,o}V_{j}a_{_{j}},
\label{Eq:Heisenberg-ak}
\end{align}
\end{subequations}
where $\kappa_{j}$ and $a_{j}^{\text{in}}$ $(j=s,o)$ are the damping rate
and noise operator of the cavities $S$ and $O$, the negative and
positive oscillators have same damping $\gamma_{m}$, and
$\Pi^{\textrm{in}} = p^{\textrm{in}}_{1} - p^{\textrm{in}}_{2}$ is
thermal noise of the oscillator, in which
$p^{\textrm{in}}_{1}~(p^{\textrm{in}}_{2})$ is noise operator of
normal~(NEM) oscillator, and the correlation function has the relation
$\langle p^{\textrm{in}}_{1} p^{\textrm{in}}_{2} \rangle =0$, then $\langle \Pi^{\textrm{in}}(t)
\Pi^{\textrm{in}}(t^\prime)\rangle = 2 \textrm{coth} \big( \frac{\hbar
\omega_{m}}{2k_{B} T} \big) \delta(t-t^{\prime})$.
From Eqs.~\eqref{Eq:Heisenberg-Q} and \eqref{Eq:Heisenberg-Pi}, it is obvious that the
collective position $Q$ and relative momentum $\Pi $ form a QBA free system.
Due to $[Q, \Pi ]=0$, the collective position $Q$ and relative momentum $\Pi $
can be simultaneously measured with arbitrary precision. The variance of $Q$~($\Pi)$ does not affect the variance of $\Pi $~($Q$) although the $\Pi $
is related with $Q$~(see Eq.~\eqref{Eq:Heisenberg-Pi}). Since the cavity $S$~($O$) works as a sensor~(detector), the dissipation should be included
because it is an open system in order to sense~(output) signal, while the
cavity chain functions as a transmitter, and it is reasonable to
ignore the loss of the chain for high quality cavities.

Through integrating Eq.~\eqref{Eq:Heisenberg-ak}, the formal solution of $a_{k}(t)$ can be obtained
\begin{equation}
\begin{split}
a_{k}(t)=& a_{k}(0)e^{-i\Delta_{k}t}-i\int_{0}^{t}\dd{\tau}e^{-i\Delta
_{k}(t-\tau)} \sum_{j=s,o} V_j a_{j}(\tau).
\end{split}
\label{Eq:solution-ak}
\end{equation}
Inserting Eq.~\eqref{Eq:solution-ak} into Eqs.~\eqref{Eq:Heisenberg-aa} and~\eqref{Eq:Heisenberg-ab}, we obtain
\begin{align}
\dot{a}_{s}=& -i(\Delta_{s}-i\frac{\kappa_{s}}{2})a_{s}-iga_{s}Q+A_{s}^{\text{in}}-\int_{0}^{t}\dd{\tau}\int \dd{\omega}\sum_{j=s,o}J_{sj}(\omega)a_{j}(\tau)e^{-i\omega (t-\tau)},  \label{Eq:a_at}  \\
\dot{a}_{o}=& -i(\Delta_{o}-i\frac{\kappa_{o}}{2})a_{o}+E_{o}+A_{o}^{\text{in}} -\int_{0}^{t}\dd{\tau}\int \dd{\omega }\sum_{j=s,o}J_{oj}(\omega)a_{j}(\tau)e^{-i\omega (t-\tau)},  \label{Eq:a_bt}
\end{align}
where $A_{j}^{\text{in}}=\widetilde{a}_{j}^{\text{in}}+\sqrt{\kappa_{j}}a_{j}^{\text{in}}$
is the noise operator, and $\widetilde{a}_{j}^{\text{in}}=
-i\sum_{k} V_j a_{k}(0)e^{-i\Delta_{k}t}$ is the noise
operator of the structured reservoir.
With the transform $\sum_{k} \to \int \dd{\omega }
\dv*{k}{\omega} = \int \dd{\omega} \varrho(\omega)$~\cite{Tan2011} in
which $\omega$ means relative frequency  $\Delta_{k} = \omega_{k} -
\omega_{d}$, that is $\omega =
\Delta_{w} - 2\xi_{w} \cos(k)$ with $\Delta_w = \omega_{w} - \omega_d$, we can obtain the
spectrum function as
\begin{equation}
J_{ij}(\omega)=\varrho (\omega)V_{i}^{* }(\omega)V_{j}(\omega),
\label{Eq:spectrum-define}
\end{equation}
where
\begin{equation}
\varrho (\omega)=\frac{1}{\sqrt{\qty(2\xi_{w})^{2}-\qty(\Delta_{w}-\omega
)^{2}}},
\label{Eq:spectrum-rho}
\end{equation}
and
\begin{equation}
V_{i}(\omega)=\sqrt{\frac{2}{\pi }}\xi_{i}\sin \qty [ n_{i}
\arcsin(\sqrt{1-\qty(\frac{\omega -\Delta_{w}}{2\xi_{w}})^{2}})].
\label{Eq:spectrum-V}
\end{equation}

Considering the sensing cavity pumping with classical field, we can expand
the cavity field as $a_{j}\rightarrow \alpha_{j}+a_{j}$~($j=s,o$), which
means that the cavity field can be decomposed to the classical mean
value $\alpha_{j}$ plus its quantum part, so that the dynamical
equation can be linearized. In strong non-Markovian regime, $\alpha_{j}$ does not mean the
steady-state values of the cavity field $a_{j}$.
From Eqs.~\eqref{Eq:Heisenberg-Q} and~\eqref{Eq:Heisenberg-Pi}, it is easy to verify that $\langle Q\rangle $ is
independent of the cavity fields due to the interference between the two
oscillator, which indicates that the self-sustained oscillation of
optomechanical system which is an obstacle of linearization is suppressed.
The situation is different from the generic optomechanical
system~\cite{Zhou2013,Zhang2016b} where the zero point of the mechanical oscillator is
displaced due to the radiation pressure. According to Eq.~\eqref{Eq:a_at}
and~\eqref{Eq:a_bt}, we have
\begin{equation}
\dot{\vb* {\alpha }}=-i\vb{\widetilde{\Delta}}\vdot\vb* {\alpha }-
\vb{E}+\int_{0}^{t}\dd{\tau}\int \dd{ \omega } \vb{J}(\omega)\vdot
\vb* {\alpha }(\tau)e^{-i\omega (t-\tau)},  \label{Eq:dt-alpha}
\end{equation}
where
\begin{equation}
\vb{J}(\omega)=
\begin{bmatrix}
J_{so}(\omega) & J_{ss}(\omega) \\
J_{oo}(\omega) & J_{os}(\omega)
\end{bmatrix} ,
\end{equation}
$\vb*{\alpha}(t)=[\alpha_{s}(t),\alpha_{o}(t)]^{\text{T}}$,
$\vb{\widetilde{\Delta}}=\text{diag}[\widetilde{\Delta }_{s},\widetilde{\Delta }_{o}]$
with $\widetilde{\Delta}_{s}=\Delta_{s}-i\frac{\kappa_{s}}{2}
+g\langle Q\rangle$, $\widetilde{\Delta}_{o}=\Delta_{o}-i\frac{\kappa
_{o}}{2}$ and $\vb{E}=[0,E_{o}]^{\text{T}}$. We perform the Laplace
transformation $O(z)=\int_{0}^{\infty }\dd tO(t)e^{izt}$ to solve the
dynamic evolution $\alpha_{j}$, with initial values
$\vb*{\alpha}(0)=0$, we can obtain
\begin{equation}
z\vb* {\alpha }(z)=\vb{\widetilde{\Delta}}\vdot\vb* {\alpha }(z)-\frac{
1}{z}\vb{E}+\vb* {\sigma }(z)\vdot\vb* {\alpha }(z),
\end{equation}
where $\vb* {\sigma }(z)=\int \dd{\omega}\frac{\vb{J}(\omega
)}{z-\omega }$ is the self-energy matrix. In order to obtain a simple and
clear meaning of the solution of above equation, we first assume $E_{o}$
absence to obtain a Green's function $\bar{\alpha}_{j}(\tau)$. Then, we
have
\begin{equation}
\bar{\alpha}_{s}(z)=i\frac{\sigma_{so}}{\mathcal{D}(z)},
\label{eq:alphabars-zm}
\end{equation}
where
\begin{equation}
\mathcal{D}(z)=[z-\widetilde{\Delta }_{s}-\sigma_{ss}(z)][z-\widetilde{
\Delta }_{o}-\sigma_{oo}(z)]-\sigma_{so}(z) \sigma_{os}(z),
\end{equation}
and for more details seeing Appendix~\ref{Sec:app-GreenFunction}. With the
relation $\alpha_{j}(t)=E_{o}\int_{0}^{t}\dd\tau \bar{\alpha}_{j}(\tau)$
and the inverse Laplace transform, we obtain the solution in the long-time
limit
\begin{equation}
\begin{split}
\alpha_{s}(t\rightarrow \infty)=& \sum_{n^{\prime }=1}^{N_{b}}\frac{iE_{o}}{\omega
_{r_{n^{\prime }}}}\mathcal{Z}_{n^{\prime }}e^{-i\omega_{r_{n^{\prime
}}}t}-E_{o}\sum_{n=1}^{N_{p}}\frac{i\mathcal{Z}_{n}}{\omega_{r_{n}}}  +I_{\text{NE}},
\end{split}
\label{Eq:alpha_infinity}
\end{equation}
where $N_b$ is the number of poles on the first Riemannian sheet, and $N_p$ is the number of poles on the first and second Riemannian sheet.
We can find all the poles of $\bar{\alpha}_{s}(z)$ denoted as
$\omega_{r_{n}}$ through solving
$\mathcal{D}(z) =0$.
The poles are classical by its position. 
It can be proof that when the poles on first Riemannian sheet where
$\Re(\omega_{r_n}) < \Delta_{w} - 2 \xi_{w}$ or $\Re(\omega_{r_{n}})
>\Delta_{w} + 2 \xi_{w}$ the poles must be on the real axis. ~(for more details seeing
Appendix~\ref{Sec:app-GreenFunction} and \ref{sec:long-time-solution}).
We denote this pole as $\omega_{r_{n^{\prime}}}$ which is real
number, the first term in
Eq.~\eqref{Eq:alpha_infinity} will be exponent oscillation term
without dissipation, and this state is called bound state.
$\mathcal{Z}_{n^{\prime }}$ is the residues of at poles $\omega
_{r_{n^{\prime }}}$. All the poles on the first and second Riemannian sheet
with residues $\mathcal{Z}_{n}$ contribute to the second terms. The last
term is non-exponential decay, and it is studied in
Appendix~\ref{sec:non_exponential_decay}.
\begin{figure}[tbph]
    \centering \includegraphics[width=0.90\linewidth]{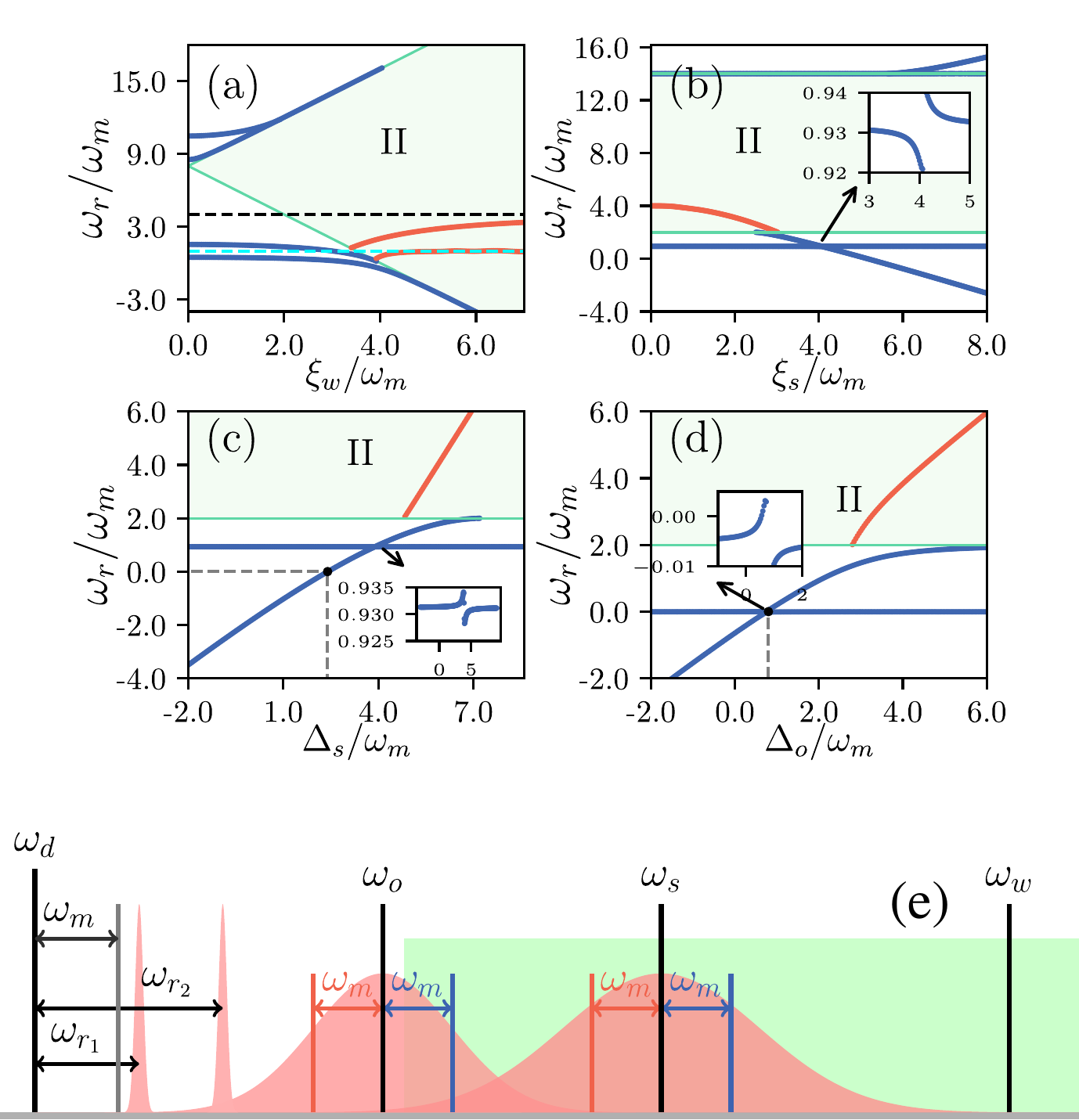}
\caption{The frequency $\protect\omega_{r}$ as the function of $\protect\xi
_{w}$~(a), $\protect\xi_{s}$~(b), $\Delta_s$~(c) and $\Delta_o$~(d); the
blue curves stand for the bound states, while the red curves for
the poles leading to exponential decay. The black and cyan dashed lines in~(a)
represent $\Delta_{s}$ and $\Delta_{o}$, respectively. The inset
of~(b), (c) and (d) for
zoom-in of the cross point.
(e) The diagram of cavity mode
and its sideband in optomechanical system, where the red peaks with
large width labeled as $\omega_o$ and $\omega_{s}$ represent the
bare cavity modes, and the green area corresponds to
$\Delta_{w} - 2\xi_{w} < \omega< \Delta_{w} + 2 \xi_{w}$~(without
bound states). The frequencies of
bound states are adjustable.
The parameters which are unspecified before, are
$\Delta_{w}=8\protect\omega_{m}$, $\Delta_{s}=4 \protect\omega_{m}$,
$\Delta_{o}=2\protect\omega_{m}$, $\protect\xi_{w}=3
\protect\omega_{m}$, $\protect\xi_{s}=4\protect\omega_{m}$,
$\protect\xi_{o}=2 \protect\omega_{m}$, $E_{o}=2\times
10^{5}\protect\omega_{m}$, $\protect\kappa_{s}=0.01\protect\omega_{m}$,
$\protect\kappa_{o}=0.05\protect\omega_{m}$
$g=0.002\protect\omega_{m}$, and $N=30$.}
\label{fig:BoundStates}
\end{figure}

Since the cavity chain is equivalent to a reservoir, the reservoir will
induce dissipation for the cavity $S$ and $O$. In order to transmit weak signal, we expect
that the cavity chain can transfer the signal without dissipation.
Fortunately, the bound state is exponent oscillator without dissipation and  can fulfil the task. In other word, bound state
is important and play the special role in the remote weak signal detection.
We now discuss the parameter region in which the bound state exists.

We numerically calculate the poles of $\bar{\alpha}_{s}(z)$ and plot the real
part of pole vs several parameters shown in Fig.~\ref{fig:BoundStates}. In
the green area, the condition $\Delta_{w}-2\xi_{w}<\omega_{r}<\Delta
_{w}+2\xi_{w}$~(determined by making positivity of the radicand of
Eq.~\eqref{Eq:spectrum-rho}) is
satisfied, which stands for the second Riemannian sheet. Beyond the green
area, we call it first Riemannian sheet. As it is shown in
Fig.~\ref{fig:BromwichPath} in the Appendix~\ref{Sec:app-GreenFunction}, the poles on the first Riemannian sheet can keep the oscillation
well, while the poles on the second Riemannian sheet can turn the
oscillation into exponential decay because of the large imaginary
part. Therefore, the poles in the first Riemannian sheet is superior than
the ones in the second Riemannian sheet in against dissipation.

The blue curves in the Fig.~\ref{fig:BoundStates}(a) represent the poles on
the first Riemannian sheet, while the red ones represent the poles on the
second sheet. With the increasing of $\xi_{w}$, the blue curves converge to
the border of the light-green area, or vanish, which means that small value
of $\xi_{w}$ is favor of surviving of bound states. As we can see from Eq.~\eqref{Eq:spectrum-define}, $\xi_{w}$ determines the width of spectrum, the
larger $\xi_{w}$, the weaker non-Markovianity. This phenomenon reflects
that with a large width of spectrum the system converge to a Markovian
system where poles converge to $\Delta_{s}$ and $\Delta_{o}$. In other
word, strong non-Markovianity does not need strong nearest-neighbor coupling
of cavities.

In Fig.~\ref{fig:BoundStates}(b), we study the poles as the function
of $\xi_{s}$.
When $\xi_{s}$ is small, we can obtain two poles with the same value
of $\xi_{s}$, but only one pole is in the first Riemannian sheet, and
the other poles is in the second Riemannian sheet. With the increasing of
the value of $\xi_{s}$, the two poles are both in first Riemannian sheet,
and a high-frequency bound state is generated. Therefore, small $\xi_{w}$
and large $\xi_{s}$ are benefit to obtain the bound states. In addition,
one of the poles almost is not affected by the value of $\xi_{s}$, only the
other pole strongly depend on the value of $\xi_{s}$, which means that the
bound states are mainly determined by the coupling between the cavity
$S$~($O$) and the common environment, respectively. However, it is
interesting to find from the inset in
Fig.~\ref{fig:BoundStates}(b) that the frequencies of bound states sharply
change when the two bound states have very close frequencies.
In this case, small disturbance can make a jump between
the two bound states.

In Fig.~\ref{fig:BoundStates}(c) and (d), we study the poles as the function
of $\Delta_{s}$ and $\Delta_{o}$. In Fig.~\ref{fig:BoundStates}(c), we
find that a bound state will disappear when $\Delta_{s} \approx 7
\omega_{m}$, and poles in second Riemannian sheet
will appear when $\Delta_{s}$ is increasing to $\Delta_{s}$ $\approx
4\omega_{m}$. To generate bound states, the frequency of cavity field
should keep away from the center frequency of CROW ($\Delta_{w}$ $=8\omega
_{m}$). The same result can be obtained for $\Delta_{o}$. The
``cross point'' in Fig.~\ref{fig:BoundStates}(c) and (d) are
similar with the one in Fig.~\ref{fig:BoundStates}(b) where
frequencies of bound states sharply change with the parameters.
In addition, we would like to mention the special points in (c) and (d) because we
will use the special points in Fig.~\ref{fig:F-s-x}. See
Fig.~\ref{fig:BoundStates}(c), when $\Delta_{s}=2.4\omega_{m}$, the
frequencies of the two bound states are $\omega_{r_{1}} \approx 0$ and
$\omega_{r_{2}} \approx \omega_{m}$. In Fig.~\ref{fig:BoundStates}(d), when
$\Delta_{s} = 2.4 \omega_{m}$, and $\Delta_{o}=0.8\omega_{m}$, we have
$\omega_{r_{1}} \approx \omega_{r_{2}} \approx 0$.

In the plotting Figs.~\ref{fig:BoundStates}(a) to (d), the
optomechanical interaction is ignored because it only appears in
$\widetilde{\Delta }_{s}=\Delta_{s}-i\frac{\kappa_{s}}{2}+g\langle
Q\rangle $, and the
displacement resulted from radiation pressure is extremely small,
compared with $\Delta_{s}-i\frac{\kappa_{s}}{2}$, so here we
choose $\langle Q\rangle =0$~\cite{M2017Quantum}. Therefore, the bound
states actually are formed by the three-body interaction among the
cavity S, O and the CROW.
In order to make the frequencies of bound state matching the frequency of the
sensing oscillator, we choose the parameters of
Fig.~\ref{fig:BoundStates} in the
unit of $\omega_{m}$ and summarize the relation of the several
frequencies of the fields in Fig.~\ref{fig:BoundStates}(e). Since the
cavity S (O) couple to a common reservoir CROW, the
effective decay rates of cavity S (O) are enhanced by the coupling
strength $\xi_{s}$~($\xi_{o}$), which is represented by the two red
peaks with large width. Fortunately, the bound states is free
of this problem. When the bound states couple to mechanical
oscillator, we can select special bound states, for instance, when
$\omega_{r_{1}}\approx \omega_{r_{2}} \approx \omega_{m}$, the sidebands are
coincident; when $\omega_{r_{1}}\approx 0$ and $\omega_{r_{2}} \approx \omega_{m}$,
one bound state resonates with driven field, and the other resonates
with oscillator; when $\omega_{r_{1}}\approx \omega_{r_{2}} \approx
0$, two bound states resonates with driven field.
We will show that in this way the output signal can be enhanced.


\section{The sensitivity of weak signal}
\label{sec:sensitivity-weakSignal}

We now study the sensitivity of weak signal detection. As we have pointed
out that the bound state means the long-life oscillation, it benefits the
transmission of the signal. Under this condition, the coupling between the
mechanical oscillator and the cavity mode $a_{s}$ is of the form
$G=\alpha_{s} g = G_{0}+\sum_{n=1}^{N_b}G_{n}e^{-i\omega_{r_{n}}t}$ where
$G_{n}=-igE_{o}\mathcal{Z}_{n}/\omega_{r_{n}}$ corresponds to bound
states $\omega_{r_{n}}$ and $G_{0} =g (I_{NE} - E_{0}
\sum_{n=1}^{N_{p}}\frac{i\mathcal{Z}_{n}}{\omega_{r_{n}}})$. The
Heisenberg-Langevin equations after linearized
can be obtained as
\begin{subequations}
\begin{align}
\dot{a}_{s}=& -i\widetilde{\Delta }_{s}^{{}}a_{s}-iGQ+A_{s}^{\text{in}
}-\int_{0}^{t}\dd{\tau}\int \dd{\omega } \sum_{j=s,o}J_{sj}(\omega)a_{j}(\tau)e^{-i\omega (t-\tau)},
\\
\dot{a}_{o}=& -i\widetilde{\Delta }_{o}a_{o}+A_{o}^{\text{in}}-\int_{0}^{t}
\dd{\tau}\int \dd{\omega } \sum_{j=s,o}J_{oj}(\omega)a_{j}(\tau)e^{-i\omega (t-\tau)}, \\
\dot{Q}=& \omega_{m}\Pi , \\
\dot{\Pi}=& -\omega_{m}Q-2f-\frac{\gamma_{m}}{2}\Pi +\sqrt{\gamma_{m}}\Pi
^{\text{in}}\label{Eq:PIt}.
\end{align}
\end{subequations}
Without back-action evading technique, the back-action force $G
a^{\dagger}_{s} + G^{*} a_{s}$ will act
on Eq.~\eqref{Eq:PIt}~\cite{Zhang2016b}, and it adds a noise channel.
The above equations can be changed into frequency domain
\begin{subequations}
\begin{align}
\omega a_{s}(\omega)=& \widetilde{\Delta }_{s}a_{s}(\omega)+\mathscr{L}[GQ]
+\sum_{j=s,o}\sigma_{sj}a_{j}(\omega)  +iA_{s}^{\text{in}}(\omega),  \label{Eq:Q_za} \\
\omega a_{o}(\omega)=& \widetilde{\Delta }_{o}a_{o}(\omega
)+\sum_{j=s,o}\sigma_{oj}a_{j}(\omega)+iA_{o}^{\text{in}}(\omega), \\
-i\omega Q(\omega)=& \omega_{m}\Pi (\omega), \\
-i\omega \Pi (\omega)=& -\omega_{m}Q(\omega)-2f(\omega)-\frac{\gamma_{m}
}{2}\Pi (\omega)  +\sqrt{\gamma_{m}}\Pi ^{\text{in}}(\omega).  \label{Eq:PI_za}
\end{align}
\end{subequations}
We let $\mathscr{L}[GQ]=\sum_{n=0}^{N_b}G_{n}Q(\omega
-\omega_{r_{n}})$, by setting $\omega_{r_{0}}=0$~( $\omega_{r_{0}}$ is
a denotation not a bound state).
According to Eq.~\eqref{Eq:Q_za} 
and~\eqref{Eq:PI_za}, we obtain
\begin{equation}
Q(\omega)=\chi_{m}(\omega)[ 2f(\omega)-\sqrt{\gamma_{m}}\Pi
^{\text{in}}(\omega) ],
\end{equation}
where $\chi_{m}(\omega)=\omega_{m}(\omega
^{2}-\omega_{m}^{2}+\frac{i}{2}\gamma
_{m}\omega)^{-1}$ is the response function. The solution of $a_{o}$ in
frequency domain can be obtained as
\begin{equation}
\begin{split}
a_{o}(\omega)=& i\bar{\alpha}_{o}(\omega)A_{o}^{\text{in}
}(\omega)+\bar{\alpha}_{s}(\omega)\Bigg[\sum_{n=0}^{N_{b}}G_{n}
Q(\omega -\omega_{r_{n}})+iA_{s}^{\text{in}}(\omega)\Bigg],
\end{split}
\label{Eq:a_o_z}
\end{equation}
in which $\bar{\alpha}_{j}(\omega)~(j = s, o)$  is Green function
$\bar{\alpha}_{j}(t)$ in frequency domain, where $\bar{\alpha}_{j}$ is
given in Eq.~\eqref{eq:alphabarA}.
The output of the observing cavity is given
by $a_{o}^{\text{out}}=a_{o}^{\text{in}}-\sqrt{\kappa_{o}}a_{o}$.
According to Eq.~\eqref{Eq:a_o_z}, the status of the oscillator is monitored by the cavity
field. Therefore, the weak signal can be read out through homodyne detection
by measuring the quadrature~\cite{Chen2013}
\begin{equation}
M=a_{o}^{\text{out}}e^{-i\theta }+a_{o}^{\text{out}\dagger }e^{i\theta }
\end{equation}
where $\theta $ is an adjustable phase.
In the frequency domain, the relation between signal and the
quadrature can be obtained
\begin{equation}
\begin{split}
M(\omega)=& \sqrt{\kappa_{o}}e^{-i\theta }\Bigg\{\frac{a_{o}^{
    \text{in}}(\omega)}{\sqrt{\kappa_{o}}}
    -i\bar{\alpha}_{o}(\omega)A_{o}^{\text{in}}(\omega)
    +i\bar{\alpha}_{s}(\omega)A_{s}^{\text{in}}(\omega) \\
      &+\bar{\alpha}_{s}(\omega)\Bigg[\sum_{n=0}^{N_{b}}G_{n}\chi_{m}
      (\omega -\omega_{r_{n}})\Big(2f(\omega -\omega_{r_{n}}) 
      -\sqrt{\gamma_{m}}\Pi ^{\text{in}}(\omega)\Big)\Bigg]\Bigg\}+H.c..
\end{split}
\label{Eq:M_total}
\end{equation}
From above equation, we can see that the response function $\chi_{m}$
and $\bar{\alpha}_{o}$ joint together to response the weak signal. If
both of them achieve their maximum values, then we can achieve the
optimized response. It is why we discuss the bound state in
Fig.~\ref{fig:BoundStates}. In
addition, due to the
introducing of a NEM oscillator, the back-action noise of the cavity
$S$ is eliminated, only the thermal noise of $\Pi^{\text{in}}$
exists, which is different from our early work~\cite{Zhang2016b} where
the additional noise contains the back-action noise proportional to
the optomechanical coupling $G$. We would like to amplify the signal
$f$ while the noise can be suppressed as low as possible. Therefore, the
level of the noise is very important in the weak signal detection. We
employ the definition of noise force~(additional
force)~\cite{Motazedifard2016} as
\begin{equation}
F_{\text{add}}(\omega)=\eval{\frac{M(\omega)}{\partial M(\omega)/\partial
f}}_{f=0}.
\end{equation}
According Eq.~\eqref{Eq:M_total}, the addition force can be obtained as
\begin{equation}
F_{\text{add}}(\omega)=F_{o}(\omega) - \frac{\sqrt{\gamma_{m}}}{2} \Pi^{\text{in}}(\omega),
\end{equation}
where
\begin{equation}
\begin{split}
F_{o}(\omega)=& \frac{\mathcal{A}(\omega)e^{-i\theta }}{\sqrt{\kappa_{o}}}
\Big\{a_{o}^{\text{in}}(\omega)+i\sqrt{\kappa_{o}}[-\bar{\alpha}_{o}(\omega)A_{o}^{\text{in}}(\omega) +\bar{\alpha}_{s}(\omega)A_{s}^{\text{in}}(\omega)]\Big\}+H.c.
\end{split}
\end{equation}
is noise induced by the cavity fields in which
\begin{equation}
\mathcal{A}(\omega)=\qty [2 \sum_{n=0}^{N_{b}}e^{-i\theta}\bar{\alpha}_{s}(\omega
)G_{n}\chi_{m}(\omega -\omega_{r_{n}})+ H.c.]^{-1}.
\end{equation}
The noise spectrum can defined as
\begin{equation}
S_{\text{add}}(\omega)=\frac{1}{2}\int \dd{\omega^{\prime}}\Big\langle F_{
\text{add}}(\omega)F_{\text{add}}(\omega ^{\prime })+(\omega
\leftrightarrow \omega ^{\prime })\Big\rangle.  \label{Eq:Sadd-define}
\end{equation}
If the signal is more weak than the noise, it definitely can not be
detected from the background noise. Therefore, the noise level
determines the accuracy of the weak signal detection.
So, it is reasonable to define the force sensitivity using
\begin{equation}
F_{s}(\omega)=\sqrt{\hbar m\omega_{m}S_{\text{add}}(\omega)},
\end{equation}
where $\hbar m\omega_{m}$ is introduced to recover the units because we
have nondimensionalized the Hamiltonian basing on $\omega_{m}$ and
$\hbar$~\cite{Ranjit2015}. The thermal noise operator $\Pi^{\text{in}}$ is the
incoherent superposition of the thermal noise of the positive and negative
oscillators.
Considering the same frequency and damping rate of positive and
negative oscillators,
the correlation function in frequency domain can be obtained as
\begin{equation}
\langle \Pi ^{\text{in}}(\omega ^{\prime })\Pi ^{\text{in}}(\omega)\rangle
=2 \coth (\frac{\hbar \omega_{m}}{2k_{B}T})\delta (\omega +\omega^{\prime }),  \label{Eq:PI-cor}
\end{equation}
where $T$ is temperature and $k_{B}$ is the Boltzmann constant~\cite{Giovannetti2001}. The noise operator of
cavity fields consist of two parts whose correlation functions are
\begin{equation}
\langle a_{i}^{\text{in}}(\omega)a_{j}^{\text{in}\dagger }(\omega ^{\prime
})\rangle =\delta_{ij}\delta (\omega +\omega ^{\prime }),  \label{Eq:a-cor}
\end{equation}
and
\begin{equation}
\langle \widetilde{a}_{i}^{\text{in}}(\omega)\widetilde{a}_{j}^{\text{in}\, \dagger
}(\omega ^{\prime })\rangle =J_{ij}(\omega)\delta (\omega +\omega ^{\prime
}).  \label{Eq:at-cor}
\end{equation}
We can obtain the additional noise spectrum as
\begin{equation}
\begin{split}
S_{\text{add}}(\omega)=& \frac{1}{2}\Bigg \{ \gamma_m \coth (\frac{\hbar \omega_{m}}{
2k_{B}T})+|\mathcal{A}(\omega)|^{2}\Bigg[\frac{1}{\kappa_{o}} +\kappa_{o}|\bar{\alpha}_{o}(\omega)|^{2}+\kappa_{s}|
\bar{\alpha}_{s}(\omega)|^{2}+\mathcal{B}(\omega)\Bigg] \\
& \quad +\omega \leftrightarrow -\omega \Bigg \},
\end{split}
\label{Eq:S-add}
\end{equation}
where
\begin{equation}
\mathcal{B}(\omega)=\vb*{\bar{\alpha}}(\omega)\vdot\vb{J}
(\omega)\vb*{\bar{\alpha}}^{\dagger }(-\omega)
\end{equation}
stands for the noise induced by the non-Markovian reservoir.
When $\omega <\Delta_{w}-2\xi_{w}$ or $\omega >\Delta_{w}+2\xi_{w}$
which is out of the definition of the spectrum, $\mathcal{B}(\omega)$
will vanish, the corresponding noise is eliminated. However, the
disappearance of $\mathcal{B}(\omega)$ does not mean the lowest noise
because the other noise in Eq.~\eqref{Eq:S-add} may be amplified when $\mathcal{B}(\omega)$ vanishes.

\begin{figure}[tbph]
    \centering \includegraphics[width=0.95\linewidth]{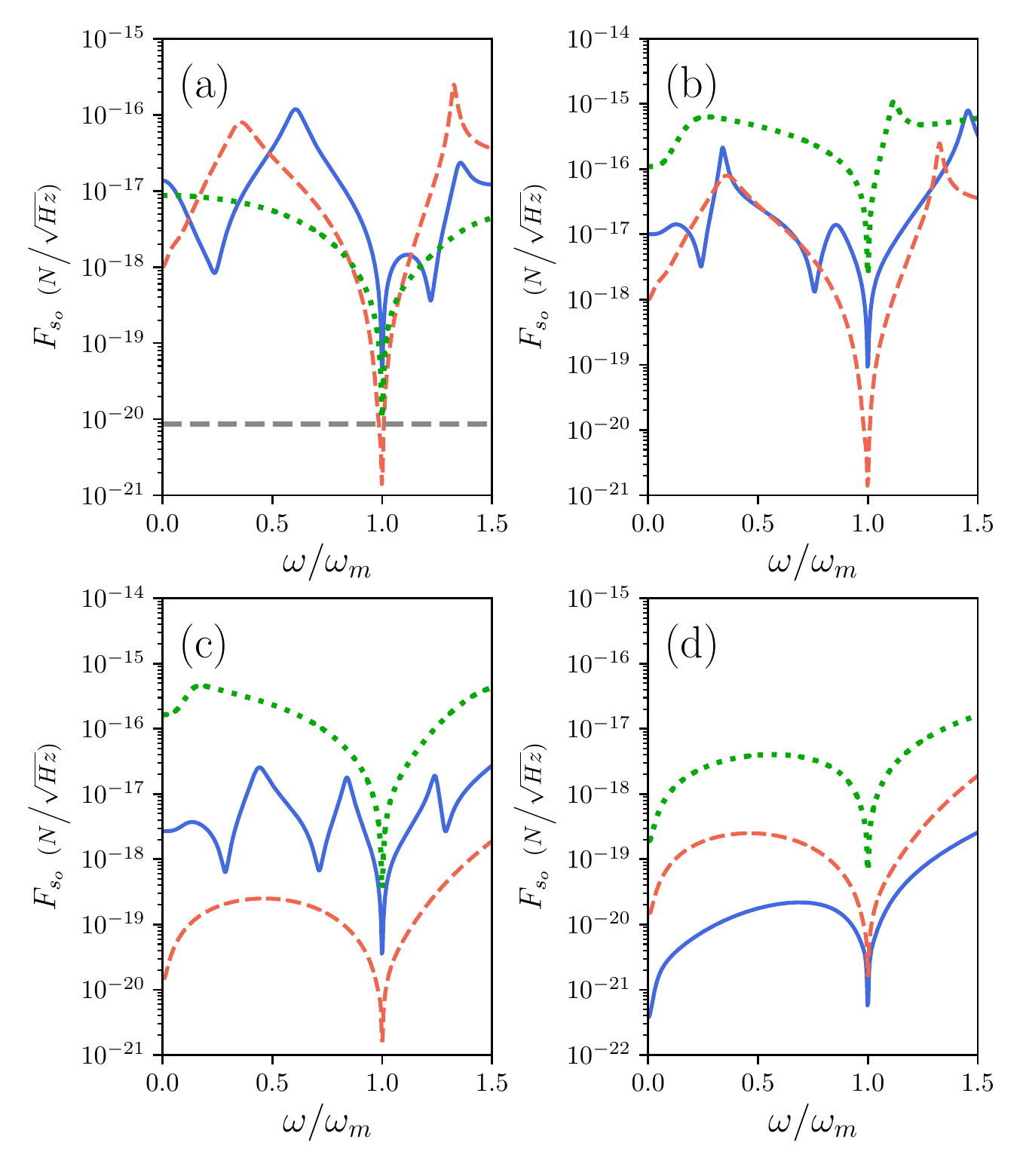}
\caption{The sensitivity of the sensor (the thermal noise absent)
    affected by several parameters where (a):
    $\protect\xi_{w}=2.4\protect\omega_{m}$~(blue solid),
    $3\protect\omega_{m}$~(red dashed) and $12\protect\omega_{m}
    $~(green dotted), (b): $\protect\xi_{s}=6\protect\omega_{m}$~(blue
    solid), $4\protect\omega_{m}$~(red dashed) and
    $8\protect\omega_{m}$~(green dotted), (c):
    $\Delta_{s}=2\protect\omega_{m}$~(blue solid),
    $2.4\protect\omega_{m}$~(red dashed), and
    $8\protect\omega_{m}$~(green dotted), and (d):
    $\Delta_{o}=0.8\protect\omega_{m}$~(blue solid),
    $2\protect\omega_{m} $~(red dashed), and
    $6\protect\omega_{m}$~(green dotted), where
    $\Delta_{s}=2.4\protect\omega_{m}$ in~(d). The grey dashed line
    in~(a) represents the sensitivity corresponding to zero-point
    fluctuation noise. Here,$\gamma_{m}=10^{-5}\omega_m$,
    $\protect\omega_{m}/2\protect\pi
    =0.5$GHz, $m=1.4 \times 10^{-18}$kg
    and $\protect\theta =\protect\pi /2$.  Other parameters which are
unspecified are same with Fig.~\ref{fig:BoundStates}.}
\label{fig:F-s-x}
\end{figure}

We now investigate the optical noise $F_{s_{o}}$ which is the
sensitivity $F_{s}$
when we temporally ignore the thermal noise of the oscillator. In
Fig.~\ref{fig:F-s-x}(a), the green curve with $\xi_{w}=12\omega_{m}$
corresponds to
no bound state, and that with $\xi_{w}=2.4\omega_{m}$, $3\omega_{m}$
correspond to the bound states. The red curve reaches to an ultrahigh
sensitivity which is smaller than the zero-point fluctuation of
oscillator~(grey dashed line), when the bound states resonate with
mechanical oscillator, i.e., around $\omega /\omega_{m}=1$.
Notice for the three values of $\xi_{w}$, the sensitivity is not
monotonously affected by $\xi_{w}$. For larger value of $\xi_{w}$,
there is no bound state but stronger field input, while
for small $\xi_{w}$, the CROW can not efficiently transfer the information.
The results corresponds to Fig.~\ref{fig:BoundStates}(a). The blue
curve as well as the red one has at least two dips. The blue curve
exhibits more sideband because the bound states do not resonate with the oscillator.

We next study the additional noise $F_{s_{o}}$ affected by $\xi_{s}$, shown in
Fig.~\ref{fig:F-s-x}(b).
Though a large $\xi_{s}$ ensures the existence of bound state, an
overlarge $\xi_{s}$ will increase the effective lose of cavity field.
This is because $\xi_{s}$ is the coupling between the cavity $S$ and the CROW~(reservoir),
meaning more photons are lost and the linearized optomechanical coupling is
decreased. Therefore, the green curve is above the red one,
in Fig.~\ref{fig:F-s-x}(b). For $\xi_{s}=4\omega_{m}$,
$\omega_{r_{1}}\approx \omega_{r_{2}}\approx \omega_{m}$ \ leads to the
linearized cavity field $a_{s}$
with the frequency $\omega_{r_{1}}$~($\omega_{r_{2}}$). Then the cavity
field will resonate to the sensing mechanical oscillator; therefore we can
achieve high sensitivity, see the red curve in Fig.~\ref{fig:F-s-x}(b).
When $\xi_{s}=6\omega_{m}$, the high sensitivity can be reached at the
sidebands, though it is
worse than the red curve where the two bound states resonate with
oscillator. When bound states do not resonate with oscillator, we can
realize a wideband detection, though it decreases the sensitivity slightly.

In Fig.~\ref{fig:F-s-x}(c), we show the effect of the detuning
$\Delta_{s}$ on the sensitivity.
For the green curve, a bound state vanishes and a pole corresponding
to exponential decay appears, which is demonstrated in
Fig.~\ref{fig:BoundStates}(c).
Therefore, the sensitivity without bound state is worse than that with
bound states~(blue and red curves).
When $\Delta_{s}=2\omega_{m}$~(corresponding to the blue curve in
Fig.~\ref{fig:F-s-x}(c)), we can find the $\omega_{r_{1}}\approx
-0.29\omega_{m}$~(relative to the frequency of pump field
$\omega_{d}$) and $\omega_{r_{2}}\approx \omega_{m}$ (see Fig.~\ref{fig:BoundStates}(c)).
Since one of frequency of the bound states is different from
$\omega_{m}$, we can observe several dips.
For $\Delta_{s}=2.4\omega_{m}$, we have $\omega_{r_{1}}\approx 0$ and $\omega_{r_{2}}\approx
\omega_{m}$ (see Fig.~\ref{fig:BoundStates}(c)).
Under this case, the red curve shows us an ultrahigh sensitivity where
in wide range of frequency the sensitivity is better than that of red
curve, as shown in Fig.~\ref{fig:F-s-x}(a) and (b).
$\omega_{r_{1}}\approx 0$ means that the bound state resonates with
driven field, the classical part of the cavity is increased.
When $\omega_{r_{1}}$ is very
close to zero, $\alpha_{s}(\infty)$ is amplified greatly, and then the
effective optomechanical coupling is enhanced because the effective
optomechanical coupling $G$ is proportional to $\alpha_{s}(\infty)$.
Meanwhile, the other bound state with $\omega_{r_{2}}\approx \omega_{m}$ is resonant
with optomechanical oscillator.
Therefore, we can achieve ultrahigh sensitivity.

In Fig.~\ref{fig:F-s-x}(d), we investigate the sensitivity with
different $\Delta_{o}$ where we choose $\Delta_{s}=2.4\omega_{m}$.
The blue curve with $\Delta_{o}=0.8\omega_{m}$ corresponds to
$\omega_{r_{1}}\approx \omega_{r_{2}}\approx 0$ which we have
mentioned in the discussion of Fig.~\ref{fig:BoundStates}(d).
In this case, although the effective optomechanical coupling can be
enhanced, the sensitivity is not so good as that shown in red curve where
$\Delta_{o}=2\omega_{m}$ leads to $\omega_{r_{1}}\approx 0$ and
$\omega_{r_{2}}\approx \omega_{m}$.
For the green curve, only one bound state exists, and the sensitivity
is decreased.

From Fig.~\ref{fig:F-s-x}, we conclude that we can obtain high sensitivity
in three cases.
The first case is $\omega_{r_1}\approx \omega_{r_2} \approx \omega_{m}$, which realizes the sideband coupling between bound states and mechanical oscillator.
However, the bound states can not be well driven, and the linearized optomechanical coupling is not effectively enhanced.
The second case is $\omega_{r_1} \approx 0$ and $\omega_{r_2} \approx \omega_m$,
which not only the linearized optomechanical coupling is enhanced,
but also the bound stat can sideband couple with oscillator.
The third case is $\omega_{r_1} \approx \omega_{r_2} \approx 0$, where
two bound states resonate with driven field.
The last case reach the best sensitivity, due to the extremely high driving efficiency and the absence of back-action noise.

\begin{figure}[htpb]
    \centering \includegraphics[width=0.9\linewidth]{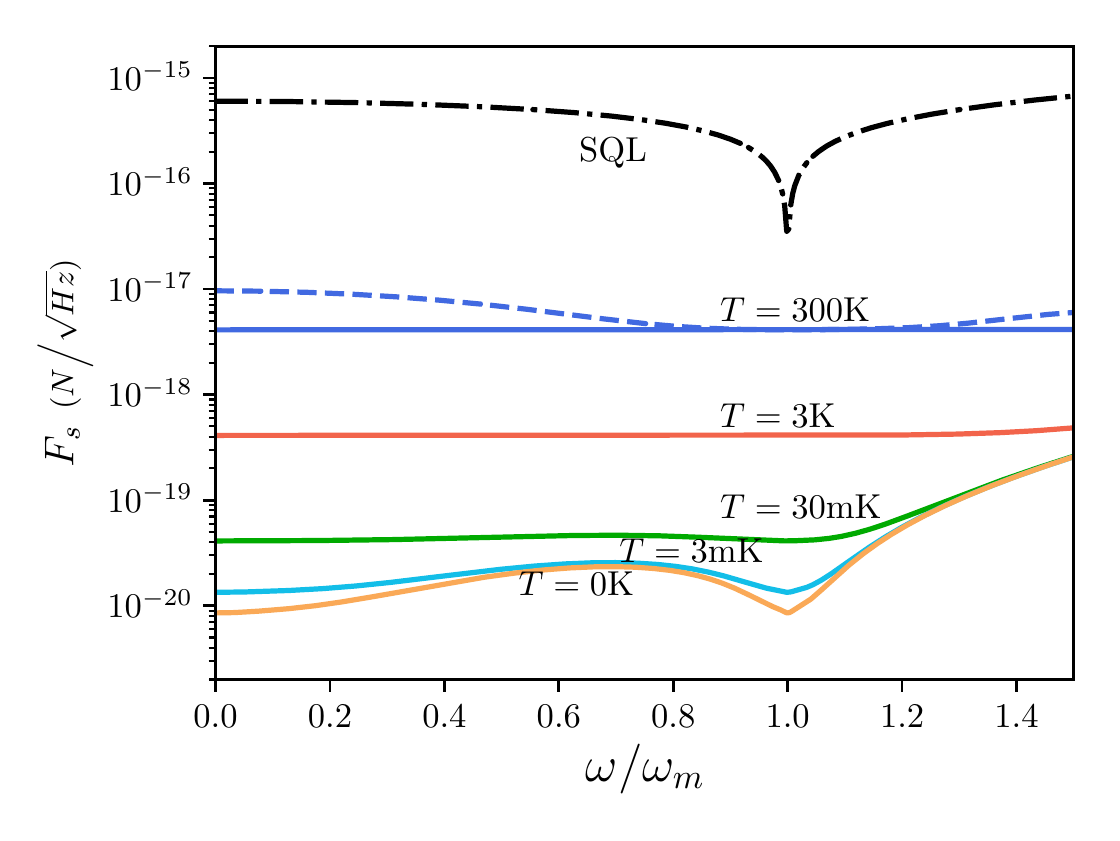}
\caption{The sensitivity for different temperature. The parameters are $\Delta_{w}=8
\protect\omega_{m}$, $\protect\xi_{w} = 3 \protect\omega_{m}$,
$\Delta_{s}=2.4\protect\omega_{m}$, $\protect\xi_{s} = 4
\protect\omega_{m}$,
$\Delta_{o} = 0.8 \protect\omega_{m}$, $\protect\xi_{o} = 2 \protect\omega_{m}$,
and $\protect\theta = \protect\pi/2$. The black dashdot line represents
the SQL of optomechanical detection where thermal noise with $T=300$K
is included. The dashed curve corresponds to
the green curve in Fig.~\ref{fig:F-s-x}(a). Other parameters which are unspecified
are the same as in Fig.~\ref{fig:BoundStates}}
\label{fig:FsWithTh}
\end{figure}

Taking the thermal noise into account, we plot the total the
sensitivity $F_{s}$ in Fig.~\ref{fig:FsWithTh}.
When detect signal in room temperature, without bound state~(blue
dashed curve), $F_{s_{o}}(\omega)$ is much larger than the thermal
noise $F_{\textrm{th}}\approx
\sqrt{m\gamma_{m}k_{B}T} \approx 4 \times 10^{-18}N/\sqrt{Hz}$ except the frequency
around $\omega_{m}$, and the optimized $F_{s_{o}}(\omega)$ in our
scheme~(blue solid curve) is limited by $F_{\textrm{th}}$.
In weak force sensing with general optomechanical system (without
back-action evasion), photon shot noise and quantum back-action leads
to the so called standard quantum
limit~(SQL),~(see~\cite{Motazedifard2016}). As shown in
Fig.~\ref{fig:FsWithTh}, the level of SQL is higher than thermal noise
of room temperature. With back-action evasion, we can see that the
optical induced noise is almost zero because at the lowest point, the
sensitivity is only limited by the thermal noise.
That is to say, our proposal is better than generic optomechanical
sensors even in room temperature.
When $T=3K$, the thermal noise dominates the sensitivity as well. We
can calculate the sensitivity
$F_{s}\approx 4 \times 10^{-19}N/\sqrt{Hz}$, shown
in Fig.~\ref{fig:FsWithTh}. When $T=30$mK, the optical noise
significantly impacts the additional noise, especially in the
high-frequency regime. Cooling to $3mK$ where
the thermal phonon number $n_{th}\approx 2.5$, the
sensitivity is very close to that in zero temperature.

%

In the optomechanical crystal regime, high quality factor of cavity
and oscillator and the strong optomechanical coupling have been
realized~\cite{Aspelmeyer2014a}.
The optomechanical systems with embedded spin have also been
reported~\cite{Cady2019}, which may be applied to on-chip
back-action-free force sensor. 
If pre-cooling the bath of oscillator~\cite{Chan2011},
the thermal noise can be suppressed,
our proposal can reach an ultrahigh sensitivity.
Even in room temperature, in wide region of frequency, our proposal is
better than generic optomechanical sensors, in which a higher
sensitivity can be achieved under same driven field, so that the
optical induced heating can be suppressed.
Our scheme provide a guideline for remotely sensing quantum
signal or weak force.

\section{conclusion}

\label{sec:discussion_and_conclusion}

In conclusion, we provide a proposal for high sensitivity remote weak
force sensor in which the hybrid optomechanical system is back-action
free through coupling to NEM oscillator. A CROW as a non-Markovian
reservoir is employed for transfer the output field from the sensing
cavity to the observing cavity. In order to non-dissipatively transfer
the weak signals, we carefully investigate the condition of bound
states. By tuning the detuning of the cavity, we can choose
optimized bound states. With the assistance of bound
states, an ultrahigh sensitivity with the optical noise smaller than
the zero-point fluctuation, can be achieved. In frequency domain, a
high sensitivity detection not only can be achieved at
$\omega \approx \omega_{m}$ but also
in a wide range of frequency. Even in room temperature, the optimized
sensitivity with bound state is much lower than that without bound
state. When the temperature is near 3mK, the sensitivity reaches
$10^{-20}N/\sqrt{Hz}$. In our investigation, we do not included the squeezing
oscillator technique and omit the loss of CROW. If we introduce the squeezing technique and take account of the loss of the CROW, the noise of CROW may be canceled by the suppressed noise with squeezing technique.

\section{Acknowledgement}

This work was supported by NSFC under Grant No.~11874099 and 11674120.

\appendix

\section{Green's function and bound state}

\label{Sec:app-GreenFunction}
In this section, we study the Green's function of the cavity field,
and the position of the poles of the Green's function.

The cavity field $\alpha_{j}(t)$ $(j=s,o)$ can
be calculated through
\begin{equation}
\alpha_{j}(t)=E_{o}\int_{0}^{t}\dd{\tau}\bar{\alpha}_{j}(\tau)
\label{Eq:alpha_bar_alpha}
\end{equation}
where the $\bar{\alpha}_{j}(\tau)$ is the corresponding Green's function.
The Green's function $\bar{\alpha}_{j}$ obeys Dyson equation
\begin{equation}
\vb*{\bar{\alpha}}(t)=-i\vb{\widetilde{\Delta}}\vb*{\bar{\alpha}}
-\int_{0}^{t}\dd{\tau}\int \dd{\omega}\vb{J}(\omega)\vb*{\bar{
\alpha}}(\tau)e^{-i\omega(t-\tau)}
\label{eq:alphabarA}
\end{equation}
where $\vb*{\bar{\alpha}}(t)=[\bar{\alpha}_{s}(t),\bar{\alpha}_{o}(t)]^{
\text{T}}$. With the Laplace transform, we can obtain
\begin{align}
\bar{\alpha}_{s}(z)=& i\frac{\sigma_{so}(z)}{
\mathcal{D}(z)}, \label{eq:alphabar-z} \\
\bar{\alpha}_{o}(z)=& i\frac{z - \widetilde{\Delta}_{s} - \sigma_{ss}(z)}{\mathcal{D}(z)},
\end{align}
where
\begin{equation}
\mathcal{D}(z)=[z-\widetilde{\Delta}_{s}-\sigma_{ss}(z)] [z-\widetilde{\Delta
}_{o}-\sigma_{oo}(z)]-\sigma_{so}(z) \sigma_{os}(z).
\label{Eq:Dz}
\end{equation}

\begin{figure}[tph]
    \includegraphics[width=0.8 \linewidth]{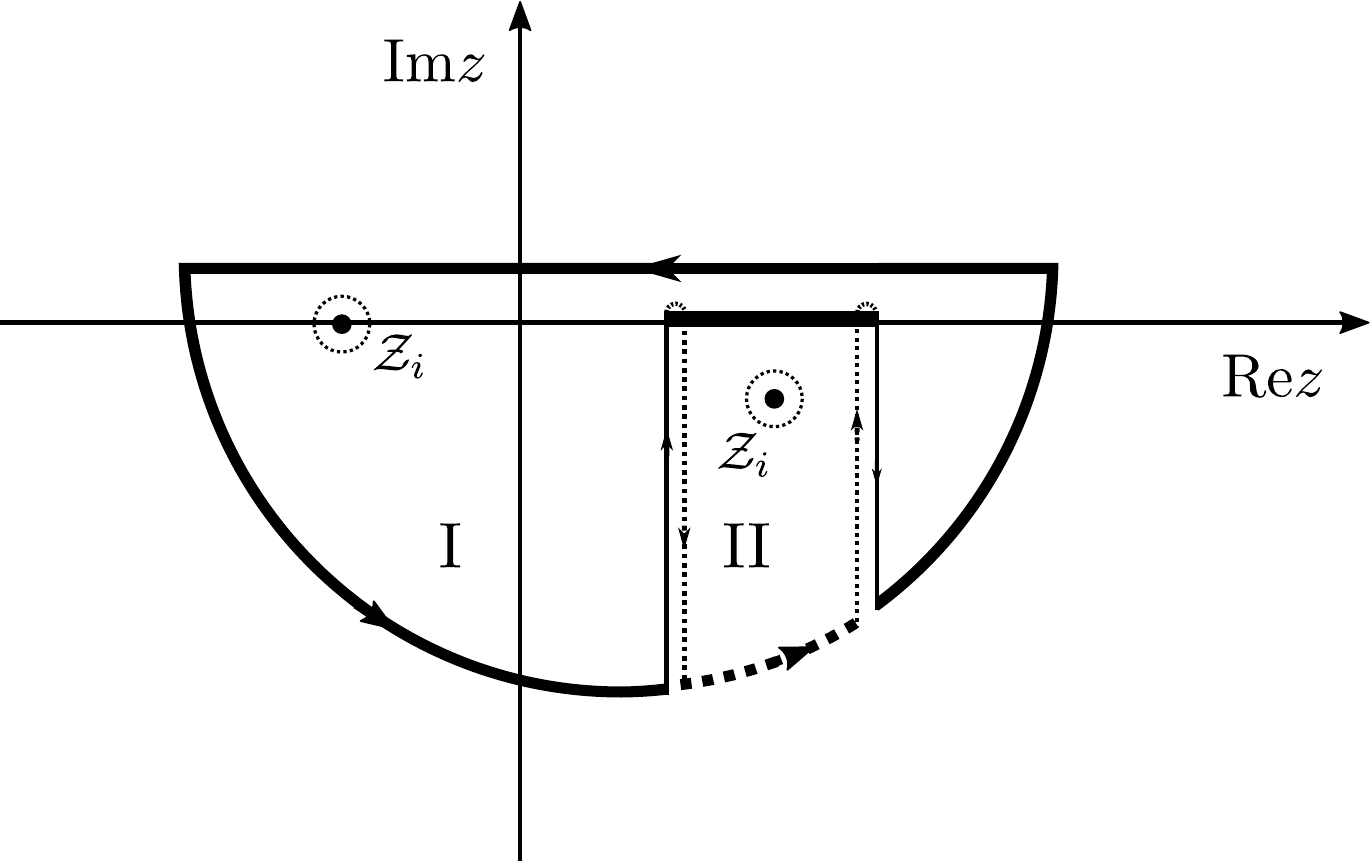}
\caption{The Bromwich path for inverse Laplace transformation}
\label{fig:BromwichPath}
\end{figure}
The $\vb*{\bar{\alpha}}(t)$ can be solved through an inverse Laplace
transformation
\begin{equation}
O(t)=\int_{i\lambda -\infty}^{i\lambda +\infty}\frac{\dd z}{2 \pi}
e^{-izt}O(z).
\end{equation}
The integral on complex plane should be calculated through contour integral
technique around the Bromwich path which is presented in Fig.\ref{fig:BromwichPath}.
For our system, $\vb*{\bar{\alpha}}(t)$ is not analytic on complex plane
where $\Delta_{w}-2\xi_{w}<z<\Delta_{w}+2\xi_{w}$ is the branch cut. The
contour path crosses the branch cut at $\Delta_{w}-2\xi_{w}$ from the first
Riemannian sheet \uppercase\expandafter{\romannumeral 1} to the second
Riemannian sheet \uppercase\expandafter{\romannumeral 2} and back from
\uppercase\expandafter{\romannumeral 2} to \uppercase\expandafter{\romannumeral 1} at $\Delta_{w}+2\xi_{w}$ for guaranteeing that the
integrand is analytic. On the second Riemannian sheet we define the analytic
continuation of $\sigma_{ij}(z)$
as the previous papers pointed~\cite{Longhi2006}.

For a CROW non-Markovian model, the poles on the
first Riemannian sheet are located on real axis ~\cite{Longhi2006}.
These poles have real part but without imaginary part that presents
the dissipation or gain, which correspond to the bound states.
The nature of bound state is that the modes corresponded the poles on the first Riemannian
sheet can not exponentially leak to the reservoir in which the
frequencies of poles are out of the cut-off frequency.
Because the CROW both couple to cavity $O$ and $S$ in our proposal, the
solution of Eq.~\eqref{Eq:Dz} is very complicated, and we can not
analytically solve it.
The mechanism of bound states is same, that is the mode with poles on
first Riemannian sheet is prevented from exponentially leaking to the reservoir.
The poles on first Riemannian sheet should be on the
real axis in our proposal. The numerical solution supports that
deduction as well.
There is another sort of poles
located on the second Riemannian sheet. These poles stand for exponential
decay with an effective decay rate and effective frequency shift, which can
be easily obtained after the Weisskopf-Wigner approximation in open quantum
system regime. When the environment spectrum tends to an infinity flat
spectrum that is to say a typical Markovian condition, the poles on the
second Riemannian sheet corresponds to the exponential decay of system.

\begin{figure}[htpb]
    \centering \includegraphics[width=0.95\linewidth]{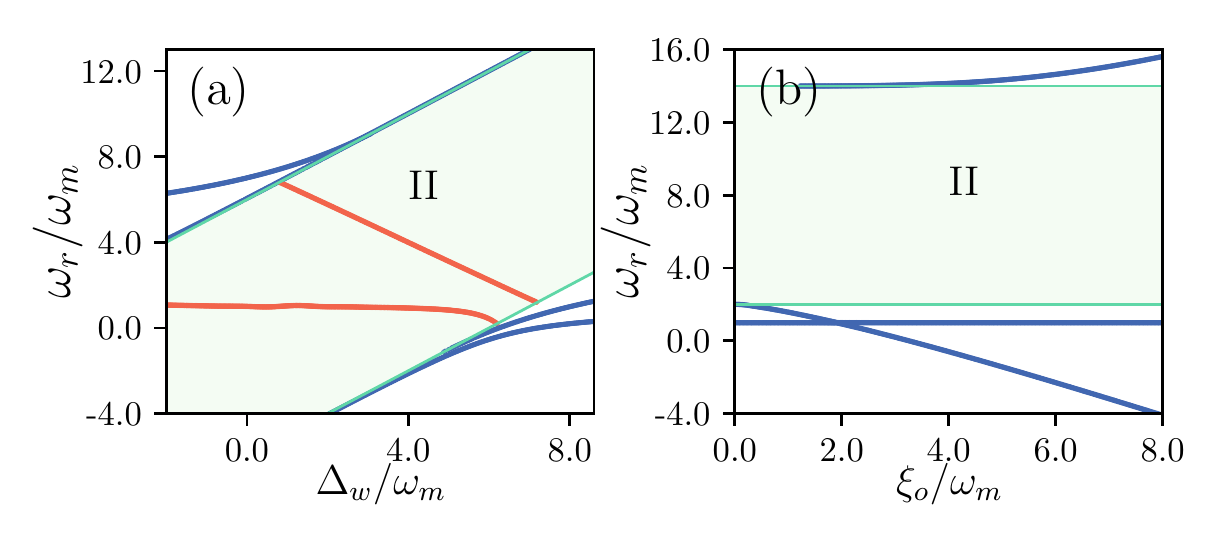}
\caption{The frequency $\protect\omega_{r}$ as the function of
    $\Delta_{w}$~(a) and $\protect\xi_{o}$~(b). Other parameters are
same with Fig.~\ref{fig:BoundStates}}
\label{fig:BS-app}
\end{figure}
We have discussed the $\omega_{r}$ as the function of $\xi_{w}$,
$\xi_{s}$, $\Delta_{s}$ and $\Delta_{o}$, in main body. As shown in
Fig.~\ref{fig:BS-app}(a), the bound state only exist when the
frequency of cavity $S$ and $O$ significantly mismatches the frequency
of the cavity of CROW. When $\Delta_{w}> 6 \omega_{m}$, there are two
bound states, and the difference of their frequency increases with
$\Delta_{w}$. With the changing of $\Delta_{w} $, the exponential
decay will disappear suddenly, while the bound states continually
change. As shown in Fig.~\ref{fig:BS-app}(b), all the poles are on the
first Riemannian sheet. For the CROW system, when $\abs{\Delta_{w}
-\Delta_{o}} \gg \xi_{w}$, it is strong non-Markovian, where the bound
states will be
generated~\cite{Longhi2006}.

\section{Long-time solution of classical mean value}

\label{sec:long-time-solution}
With the inverse Laplace transform of Eq.~\eqref{eq:alphabar-z} and
contour integral technique, we obtain
\begin{equation}  \label{Eq:bar_alpha_t}
\begin{split}
\bar{\alpha}_{s}(t)=&\sum_{n}\mathcal{Z}_{n}e^{-i\omega_{r_n}}+\qty(
\int_{\omega_{1}-i\infty}^{\omega_{1}}-\int_{\omega_{2}-i\infty
}^{\omega_{2}})\frac{\dd z}{2 \pi} e^{-izt} \qty[ \bar{\alpha}_{s}^{\text{\RomanN{2}}}(z)
-\bar{\alpha}_{s}^{\text{\RomanN{2}}}(z)],
\end{split}
\end{equation}
where $\mathcal{Z}_{n}$ is the residues of $\bar{\alpha}_{s}(z)$
corresponding to the pole at $z=\omega_{r_n}$ on the first and the second
Riemannian sheet, and the last term describes the non-exponential decay.

According to Eqs.~\eqref{Eq:bar_alpha_t} and~\eqref{Eq:alpha_bar_alpha}, we
can obtain $\alpha_s(t)$. Especially, we focus on the long-time limit $
\alpha_{s}(t \to \infty)$ which plays a key role in our proposal. The fast
decay term of $\alpha_s(t)$ are negligible, thus we can obtain
\begin{equation*}
\alpha_{s}(t\rightarrow \infty)=\sum_{n^\prime}\frac{iE}{\omega
_{r_{n^\prime}}}\mathcal{Z}_{n^\prime}e^{-i\omega_{r_{n^\prime}}t}- i E
\sum_{n} \frac{\mathcal{Z}_n}{\omega_{r_n}} + I_{\text{NE}},
\end{equation*}
where only the poles corresponding to bound state contribute to the first
term with notation $n^\prime$, the second term takes all the poles into
consideration with notation $n$, and the last term
\begin{equation}
\begin{split}
I_{\text{NE}} =& E_s \int_{0}^{t} \dd{\tau} \qty(
\int_{\omega_{1}-i\infty}^{\omega_{1}}-\int_{\omega_{2}-i\infty
}^{\omega_{2}})\frac{\dd z}{2 \pi}  e^{-iz \tau} \qty[\bar{\alpha}_{s}^{\text{\RomanN{1}}}(z)-
\bar{\alpha}_{s}^{\text{\RomanN{2}}}(z)],
\end{split}
\end{equation}
is contributed by the non-exponential decay.

The bound states may exist or not. If no bound states $G=\alpha_{s}g$ is
time-independent, otherwise $G=G_{0}+\sum_{j}G_{r_{j}}e^{- \omega_{r_{j}}t}$.

\section{non-exponential decay}

\label{sec:non_exponential_decay}
In the non-Markovian system, the non-exponential decay appears
frequently, which results from the violation of Weisskopf-Wigner
approximation. Here, it reflects in the second term of
Eq.~\eqref{Eq:bar_alpha_t} which is the integral
through vertical path, as shown in Fig.~\ref{fig:BromwichPath}:
\begin{equation}
\bar{I}_1(t)= \int_{\omega_{1}-i\infty}^{\omega_{1}}\frac{\dd z}{2 \pi}
e^{-izt} \qty[\bar{\alpha}_{s}^{\textrm{\RomanN{1}}}(z)-
\bar{\alpha}_{s}^{\textrm{\RomanN{2}}}(z)],
\end{equation}
with the transformation $x=i(z-\omega_{1})$, we obtain
\begin{equation}
\bar{I}_1(t)= \frac{ie^{-i\omega_{1}t}}{2 \pi} \int_{0}^{\infty} \dd{x}
e^{-xt} \qty[\bar{\alpha}_{s}^{\textrm{\RomanN{1}}}(\omega
_{1}-ix)-\bar{\alpha}_{s}^{\textrm{\RomanN{2}}}(\omega_{1}-ix)],
\label{Eq:Ibar1}
\end{equation}
The third term of Eq.~\eqref{Eq:bar_alpha_t} can be obtained by similar
operation as
\begin{equation}
\bar{I}_2(t)= - \frac{ie^{-i\omega_{2}t}}{2 \pi} \int_{0}^{\infty} \dd{x}
e^{-xt} \qty[\bar{\alpha}_{s}^{\textrm{\RomanN{1}}}(\omega
_{2}-ix)-\bar{\alpha}_{s}^{\textrm{\RomanN{2}}}(\omega_{2}-ix)].
\label{Eq:Ibar2}
\end{equation}
In general, $\bar{I}_1(t)$ and $\bar{I}_2(t)$ indicate the
non-exponential decay in non-Markovian dynamic. As the
Eqs.~\eqref{Eq:Ibar1} and~\eqref{Eq:Ibar2} indicating, the two
integrals are convergent. The influence of non-exponential decay can
be taken into consideration through numerical calculating the
$\bar{I}_1(t)$ and $\bar{I}_2(t)$. In general, the non-exponential
decay can decrease the loss of photon number of sensing cavity,
comparing the exponential decay.
%
\end{document}